\documentclass[modern]{aastex631}

\usepackage{amsmath}
\usepackage{subfigure}
\usepackage{comment}
\newcommand{\cca}{Center for Computational Astrophysics, Flatiron Institute, 162 Fifth Avenue, New York, NY 10010, USA}
\newcommand{\sbu}{Department of Physics and Astronomy, Stony Brook University, Stony Brook, NY 11794, USA}
\newcommand{\pa}{Department of Physics and Astronomy, Northwestern University, 2145 Sheridan RD, Evanston, IL 60208, USA}
\newcommand{\ciera}{Center for Interdisciplinary Exploration and Research in Astrophysics (CIERA), Northwestern University, 1800 Sherman Ave, Evanston, IL 60201, USA}
\newcommand{\skai}{The NSF-Simons AI Institute for the Sky (NSF-Simons SkAI), 875 N. Michigan Ave., Suite 4010, Chicago, IL 60611}

\usepackage[draft,commandnameprefix=always,commentmarkup=uwave]{changes}
\definechangesauthor[name="Will", color=cyan]{W}
\definechangesauthor[name="Anarya", color=magenta]{A}
\definechangesauthor[name="Vicky", color=yellow]{V}
 
\newcommand{\alphatwoplgplp}{\ensuremath{-2.5^{+1.4}_{-3.4}}}

\newcommand{\mchigh}{17.41}
\newcommand{\mchighunits}{\ensuremath{\mchigh \, M_\odot}}
\newcommand{\nevts}{25}
\newcommand{\dNlogmpeakincludingnew}{\ensuremath{96^{+115}_{-54}}}
\newcommand{\dNlogmpeakincludingnewunits}{\ensuremath{\dNlogmpeakincludingnew \, \mathrm{Gpc}^{-3} \, \mathrm{yr}^{-1}}}
\newcommand{\dNlogmmlowincludingnew}{\ensuremath{19^{+30}_{-12}}}
\newcommand{\dNlogmmlowincludingnewunits}{\ensuremath{\dNlogmmlowincludingnew \, \mathrm{Gpc}^{-3} \, \mathrm{yr}^{-1}}}
\newcommand{\monepctplgplpincludingnew}{\ensuremath{3.13^{+0.18}_{-0.04}}}
\newcommand{\monepctplgplpincludingnewunits}{\ensuremath{\monepctplgplpincludingnew \, M_\odot}}
\newcommand{\mpeakplgplpincludingnew}{\ensuremath{9.32^{+0.73}_{-0.87}}}
\newcommand{\mpeakplgplpincludingnewunits}{\ensuremath{\mpeakplgplpincludingnew \, M_\odot}}

\newcommand{\dd}{\ensuremath{\mathrm{d}}}

\begin{document}
\title{Hiding Out at the Low End: No Gap and a Peak in the Black-Hole Mass Spectrum }
\author[0000-0002-7322-4748]{Anarya Ray}
\email{anarya.ray@northwestern.edu}
\affiliation{\ciera}
\affiliation{\skai}
\author[0000-0003-1540-8562]{Will M. Farr}
\email{will.farr@stonybrook.edu}
\email{wfarr@flatironinstitute.org}
\affiliation{\sbu}
\affiliation{\cca}

\author[0000-0001-9236-5469]{Vassiliki Kalogera}
\email{vicky@northwestern.edu}
\affiliation{\ciera}
\affiliation{\skai}
\affiliation{\pa}

\begin{abstract}
    In recent years, the existence of a gap in the mass spectrum of compact
    objects formed from stellar collapse, between the heaviest neutron stars and
    the lightest black holes, has been a matter of significant debate. The
    presence or absence of a mass gap has implications for the supernova
    mechanism, as well as being a fundamental property of the compact object
    mass function. In X-ray binaries containing black holes a gap is observed,
    but it is not known whether this is representative of a true gap in the mass
    function or due to selection effects or systematic biases in mass
    estimation. Binary black hole mergers detected from gravitational waves in
    the GWTC-3 transient catalog furnish a large sample of several tens of
    low-mass black holes with a well-understood selection function.  Here we
    analyze the \nevts{} GWTC-3 merger events (along with
    GW230529\_181500) with at least one black hole ($3 \, M_\odot < m_1$) and chirp
	masses below those of a $20\,M_\odot$--$20\,M_\odot$ merger ($\mathcal{M} <
    \mchighunits$) to uncover the structure of the low-mass black hole mass
    function. Using flexible parameterized models for the mass function, we
    find, similar to existing studies, a sharp peak in the mass function at $m
    \simeq (8-10 M_{\odot})$. We observe a steady decline in the merger rate to
    lower masses, but by less than an order of magnitude in total, and find that the first
    percentile of black hole masses in our most flexible model is $m_{1\%} =
    \monepctplgplpincludingnewunits$. In other words, this sample of low-mass
    black holes is not consistent with the existence of a mass gap.
\end{abstract}

\section{Introduction}
\label{sec:intro}

The existence of a gap in the compact object mass-spectrum between the lightest black holes and the heaviest neutron stars remains an open question with far-reaching implications. Observational evidence for the presence or absence of a \textit{lower-mass gap} in the population of compact objects formed from stellar collapse can offer new insights into the universal properties of dense matter and the uncertain mechanisms of supernova explosions~\citep{Fryer:1999ht, Fryer:2011cx, Mandel:2020qwb, Zevin:2020gma, Liu:2020uba, Patton:2021gwh, Siegel:2022gwc}. In addition, as a fundamental property of the compact object mass spectrum, the lower-mass gap will likely play a vital role in cosmological explorations using gravitational wave~(GW) observations from future detectors~\citep{Ezquiaga:2022zkx}.

Several studies relying on electromagnetic observations of low-mass X-ray binaries~(LMXBs) have reported the existence of the lower mass gap~\citep{Bailyn:1997xt, Ozel:2010su, Farr:2010tu}. They have shown that Bayesian inference of phenomenological population models for compact object masses yields a posterior distribution for minimum black-hole mass, which can be constrained given data. Using growing samples of LMXB observations and a variety of mass-distribution models, they have found with 90\% posterior probability that the minimum BH mass is $\simeq 4.5M_{\odot}$ or higher, which suggests the existence of a gap between the heaviest possible non-rotating NSs~\citep[$2-3M_{\odot}$,][]{PhysRevLett.32.324, Kalogera:1996ci, Mueller:1996pm, Ozel:2016oaf, Margalit_2017, Ai:2019rre, Shao:2020bzt, Raaijmakers:2021uju} and the lightest observed BHs. If true, these findings are indicative of the observational preference of rapid supernova instability time scales~\citep[$\sim 10 ms$,][]{Fryer:2011cx, Belczynski:2011bn, Fryer:2022lla, Siegel:2022gwc} 

However, the analyses above are susceptible to several systematic biases, which we summarize as follows. Firstly, various assumptions underlying emission models of accretion flows can significantly bias the measured BH masses, which can in turn affect the conclusions about a gap~\citep{Kreidberg:2012ud}. On the other hand, substantial selection biases can be incurred through the incorrect assumption that every draw from their mass distribution model is equally likely to be observable~\citep{Farr:2010tu, Siegel:2022gwc}. Relying on a forward modeling approach towards LMXB formation, a recent study by \cite{Siegel:2022gwc} has shown that for supernova mechanisms capable of producing BHs in the gap, transient LMXB selection effects can significantly bias the observable sample~\citep[see also, for example, ][ for similar effects of selection biases in high mass X-ray binary populations]{Liotine2023}. Hence, it is unclear whether or not existing LMXB-based measurements of the minimum BH mass can be considered unbiased observational evidence of an existent lower mass gap. The difficulty in modeling transient LMXB selection effects, and the possibility of systematic errors in single source mass measurements, therefore necessitates alternative probes of the compact object mass spectrum to ascertain the existence of a lower mass gap.

Observations of GWs from compact binary coalescences~(CBCs) by the LIGO-Virgo-KAGRA~\citep[LVK, ][]{LIGOScientific:2014pky, VIRGO:2014yos, KAGRA:2020agh} detector network offer a novel way to probe the existence of a lower mass gap that is free of such systematic biases~\citep{Farah:2021qom,  LIGOScientific:2024elc, KAGRA:2021duu}. Selection effects in GW observations are well understood and easily modeled through the recovery of simulated sources injected into detector noise realizations~\citep{Thrane:2018qnx,Mandel:2018mve,popgw3}, which can then be used to obtain the astrophysical mass distribution of compact objects as compared to the observed distributions inferred by LMXB investigations. Previous studies have attempted this exploration by analyzing all CBC observations~(including NS containing events) using a mass distribution model that allows for a shallow gap~\citep{Farah:2021qom,  KAGRA:2021duu, LIGOScientific:2024elc}. The location, depth, and width of the gap were inferred a posteriori, and Bayesian evidences were computed for values of these parameters that correspond to the presence and absence of a lower gap. 

\cite{Farah:2021qom} have found from the second gravitational wave transient catalog~\citep{LIGOScientific:2020ibl} that the parameters controlling the gap-like feature built into their mass distribution model have a posteriori values that prefer the existence of a lower mass gap with a Bayes factor of 55.0 over its absence. Subsequent studies by \cite{KAGRA:2021duu} and \cite{LIGOScientific:2024elc} have updated this number to lower values by including events from the third gravitational wave transient catalog~\citep[GWTC-3,][]{KAGRA:2021vkt} and the mass-gap event GW230529\_181500~\cite[a compact binary merger whose most massive component lies in the $3-5M_{\odot}$ mass range,][]{LIGOScientific:2024elc}. \cite{KAGRA:2021duu} have further reported that there is no significant observational preference in GWTC-3 for or against models that exhibit or rule out a lower mass gap as compared to a default mass distribution model, which is agnostic of any gap-like feature in the relevant mass range.

It is to be noted, however, that in addition to the difference in the treatment of selection biases, existing GW-based studies have explored a fundamentally different property of the compact object mass spectrum than the LMXB investigations, on which we elaborate as follows. The LMXB dataset comprises of observed systems whose only compact object components are BHs~(more massive than the heaviest possible non-rotating NS), with the corresponding analyses attempting to constrain the minimum mass of BHs either as a cutoff or percentile of the inferred BH mass function~\citep{Bailyn:1997xt, Ozel:2010su, Farr:2010tu}. The GW analyses have instead tried to model and identify a dearth of binary mergers in the full compact object mass spectrum, near the maximum NS mass, using a dataset of observed systems that comprise both BHs and/or NSs~\citep[i.e. including BNS systems,][]{Farah:2021qom, KAGRA:2021duu, LIGOScientific:2024elc}. \textit{Hence, direct comparison of existing GW and LMXB-based probes of the lower mass gap might not be fully justified given the fundamental difference between these approaches in their definitions of the lower-mass gap as a property of the compact object mass spectrum.} 

In this paper, using flexibly parametrized population models, we analyze GW
observations of CBCs whose BH components are more massive than the heaviest
possible NSs and constrain the minimum BH mass as a percentile of the inferred
BH mass function~\citep[similar to the approach of X-ray binary studies,
][]{Farr:2010tu}.  By relying on a mass cut~\citep{Roy:2025ktr} to prevent NS components from
contaminating our analysis and using a generic model for the BH mass spectrum,
we constrain the first percentile of the BH mass function to be $\sim
3M_{\odot}$ at $90\%$ confidence, given data from GWCT3 and GW230529\_181500. Our models
respond to the overall abundance of mergers with black holes in the mass range
just above that of the heaviest neutron stars, and not to any explicitly modeled
gap-like feature in the distribution functions. We show that our results are
robust against reasonable variations of the mass-cut imposed during event
selection and of the functional forms of the population models used in the
analysis. The discrepancy between our measurements of the minimum BH mass and
that of LMXB studies can therefore be attributed either to the differences in
the treatment of selection effects, systematic biases in LMXB mass measurements, or potentially to that in the evolutionary
pathways of LMXBs and BH containing CBC systems, or to systematic mis-estimation
of masses in at least some LMXB systems.

This paper is organized as follows. In Sec.~\ref{sec:methods}, we describe our population model, analysis framework, and event selection method. In Sec.~\ref{sec:results} we show our main results and discuss potential sources of systematics. In Sec.~\ref{sec:discussion} we discuss the implications of our study and then conclude in Sec.~\ref{sec:conclusion} with as summary of the main results and the scope of potential follow-up investigations.

\section{Methods}
\label{sec:methods}
\noindent
We construct flexible population models for low-mass BBHs and analyze a sample of 25 GWTC-3 observations, both with and without GW230529\_181500, to investigate features in the mass-function and assess whether or not a gap is necessary to fit the observed dataset. The population inference and event selection methods used to carry out this exploration are delineated as follows.
\subsection{Population Model}
\noindent
We model the BH mass function as a part of the compact object mass spectrum
above the maximum possible NS mass~($m_\mathrm{low}$) which we fix throughout
the analysis. To explore the low-mass BH population, we define our mass
functions as follows:
\begin{equation}
    \label{eq:intensity-definition}
    m_1 m_2 \frac{\dd N}{\dd m_1 \dd m_2 \dd V \dd t} = R(z) f\left( m_1 \right) f\left( m_2 \right) g(m_1, m_2).
\end{equation}
\noindent
The function $f$ represents the ``common'' mass function for both components in
a binary, and includes a contribution to account for neutron stars,
$f_\mathrm{NS}$, with masses below $m_\mathrm{low}$; and a contribution for
black holes with masses above $m_\mathrm{low}$, $f_\mathrm{BH}$:
\begin{equation}
    f(m) = \begin{cases}
        f_\mathrm{NS}(m) & m < m_\mathrm{low} \\
        f_\mathrm{BH}(m) & m_\mathrm{low} < m.
    \end{cases}
\end{equation}
The neutron star component is modeled with a Gaussian shape (a reasonable
approximation to the mass function of binary neutron stars observed in our
Galaxy \citep{Farr2020,Alsing2018}), with a peak at $\mu_\mathrm{NS}$, a width
$\sigma_\mathrm{NS}$, and a rate density at $m = \mu_\mathrm{NS}$ (relative to
the black hole component; see below) of $r_\mathrm{NS}$:
\begin{equation}
    f_\mathrm{NS}(m) = r_\mathrm{NS} \exp\left( - \frac{\left( m - \mu_\mathrm{NS} \right)^2}{2 \sigma_\mathrm{NS}^2} \right).
\end{equation}
The black hole component is either a broken power law with break mass $m_b$ and
power law indices $\alpha_1$ and $\alpha_2$ below and above the break:
\begin{equation}
    f_\mathrm{BH}(m) = \begin{cases}
        \left( \frac{m}{m_b} \right)^{\alpha_1} & m < m_b \\
        \left( \frac{m}{m_b} \right)^{\alpha_2} & m_b < m
    \end{cases}, 
\end{equation}
or a sum of a broken power law and a Gaussian shape peaking at the break mass
(each contributing a fraction $1-f_g$ and $f_g$ to the rate density at the break
/ peak mass):
\begin{equation}
    f_\mathrm{BH}(m) = f_g \exp\left( - \frac{\left( m - \mu \right)^2}{2 \sigma^2} \right) + \left( 1 - f_g \right) 
    \begin{cases}
        \left( \frac{m}{\mu} \right)^{\alpha_1} & m < \mu \\
        \left( \frac{m}{\mu} \right)^{\alpha_2} & \mu < m
    \end{cases}.
\end{equation}
In both cases, $f_\mathrm{BH}(m) = 1$ when $m = m_b$ or $m = \mu$, supporting
that $r_\mathrm{NS}$ is the merger rate denisty at the peak of the neutron star
component \emph{relative} to the peak of the black hole component.

The ``pairing function'' \citep{Fishbach2020} $g$ is a power law in the total
mass,
\begin{equation}
    g(m_1, m_2) = \left( \frac{1 + \frac{m_2}{m_1}}{2} \right)^{\beta}.
\end{equation}
Note that this choice of pairing function, in contrast to a power-law pairing
\citep{KAGRA:2021duu}, implies that for $\beta \neq 0$ the mass function is not
\emph{separable} into a function of $m_1$ and a function of $m_2$.  We also
explored Gaussian pairing functions,
\begin{equation}
    g\left(m_1, m_2 \right) \propto \exp\left( - \left( m_2/m_1 - \mu_q \right)^2 /
\left( 2 \sigma_q^2 \right) \right);
\end{equation} 
but found no qualitative and few quantitative differences with the results
reported here.

The merger rate evolves in redshift according to $R(z)$ with 
\begin{equation}
    R(z) = R_0 \left( 1 + \left( \frac{1}{1 + z_p} \right)^\kappa \right) \frac{\left( 1 + z \right)^\lambda}{1 + \left( \frac{1+z}{1+z_p} \right)^\kappa}, 
\end{equation}
with fixed $z_p = 1.9$, $\kappa = 5.6$, and $\lambda = 2.7$, following
\citet{Madau2014}.  For a low-mass sample like we analyze here, it is
appropriate to fix the redshift evolution of the merger rate, since such mergers
are only observable to low redshift, giving a small ``lever arm'' to constrain
the redshift evolution.

With these definitions, $R_0$ is the merger rate per natural log mass squared,
per comoving volume, per time at redshift $z = 0$, $m_1 = m_2 = \mu, m_b$.  We
use the convention $m_2 \leq m_1$.  The parameter $\alpha_1$ is the power law
index of the common mass function below (power law) or well below (i.e.\ several
$\sigma$; power law plus Gaussian) the peak of the BH mass function, and
$\alpha_2$ the power law slope above.  Our population distribution is completely
determined by 8 \textit{hyperparameters} for the broken-power-law~(BPL) model
$\vec{\lambda}=(\mu_{NS}, \sigma_{NS}, r_{NS}, \alpha_1,\alpha_2,m_b, \beta,
R_0)$, with the broken-power-law+Gaussian~(BPLG) model requiring three
additional quantities $f_g,\mu,\sigma$.

Given measurements of these hyper-parameters, the ``common'' mass function $f$ can be normalized to obtain a probability distribution
for $m_\mathrm{low} \leq m \leq m_\mathrm{high}$ which can be written as:
\begin{equation}
    \label{eq:pm-definition}
    p(m) \equiv \frac{f(m)}{\int_{m_\mathrm{low}}^{m_\mathrm{high}} \dd m' \, f(m')}.
\end{equation}
\noindent Denoting $m_{1\%}$ to be the first percentile of this distribution, we can rely on its measurements to infer the minimum BH mass and hence the existence of a lower mass-gap~\citep{Farr:2010tu}. 

\subsection{Bayesian Hierarchical Inference}
\noindent

To infer the hyperparameters characterizing our mass function models, we employ
Bayesian hierarchical methods on our selection of GWTC-3 events (described
below). Merging BBHs constitute an inhomogeneous Poisson
process.  Modeling such a process subject to observations with
measurement uncertainty and selection effects is described
\citet{Mandel:2018mve,pop-vitale,popgw2,popgw3}.
Note that here we do not model the distribution of CBC spin parameters which
amounts to fixing the same to the uninformative priors used during single event
PE, which are isotropically oriented component spins distributed uniformly in
magnitude \citep{KAGRA:2021duu}.  

We impose uniform priors on all hyperparameters. We use Hamiltonian Monte-Carlo
methods, specifically the No U Turn Sampler~\citep[NUTS, ][]{HMC, HMC-NUTS}, to
sample from the population model posterior. To ensure quick and efficient
computation of the selection function normalization \citep{Mandel:2018mve}, we
reweight the detectable sample of simulated events provided by the LVK
\citep{KAGRA:2021duu} to a base line population model that corresponds to a
fiducial value of the hyperparameters.  We monitor convergence of the
Monte-Carlo sums used to compute the selection normalization and average over
single-event likelihoods \citep{Pdet1-Farr,Pdet2-essick,Talbot2023}

\subsection{Mass cuts and Event Selection}
With the primary goal of constraining the minimum BH mass, we chose our dataset
of low mass BH that are confidently detectied (false alarm rate from at least
one LVK pipeline lower than $1 \, \mathrm{yr}^{-1}$ \citet{KAGRA:2021duu})
containing GW events based on the following mass-cuts:
\begin{equation}
	m_\mathrm{low}\leq m_1 ~\&~\mathcal{M}\leq \mathcal{M}_{\mathrm{max}}.
\end{equation}
While the lower cut on primary mass is necessary to ensure that we are modeling
features of the BH mass function, the upper cuts on chirp mass are
imposed to remove events that are not expected to contribute to the inference of
the low-mass BBH population. Our canonical analysis uses $(3M_{\odot},
\mchighunits)$ respectively for ($m_\mathrm{low},
M_{c,\mathrm{max}}$). Out of these bounds we expect $m_\mathrm{low}$ to
have the most significant impact on the inference of $m_{1\%}$. We study the
systematic changes in $m_{1\%}$ due to variations in $m_\mathrm{low}$ by re-running the analysis with different choices of $m_\mathrm{low}$ values near $3M_{\odot}$ and comparing the
corresponding $m_{1\%}$ posteriors.

\begin{figure}[htp]
    \includegraphics[width=\columnwidth]{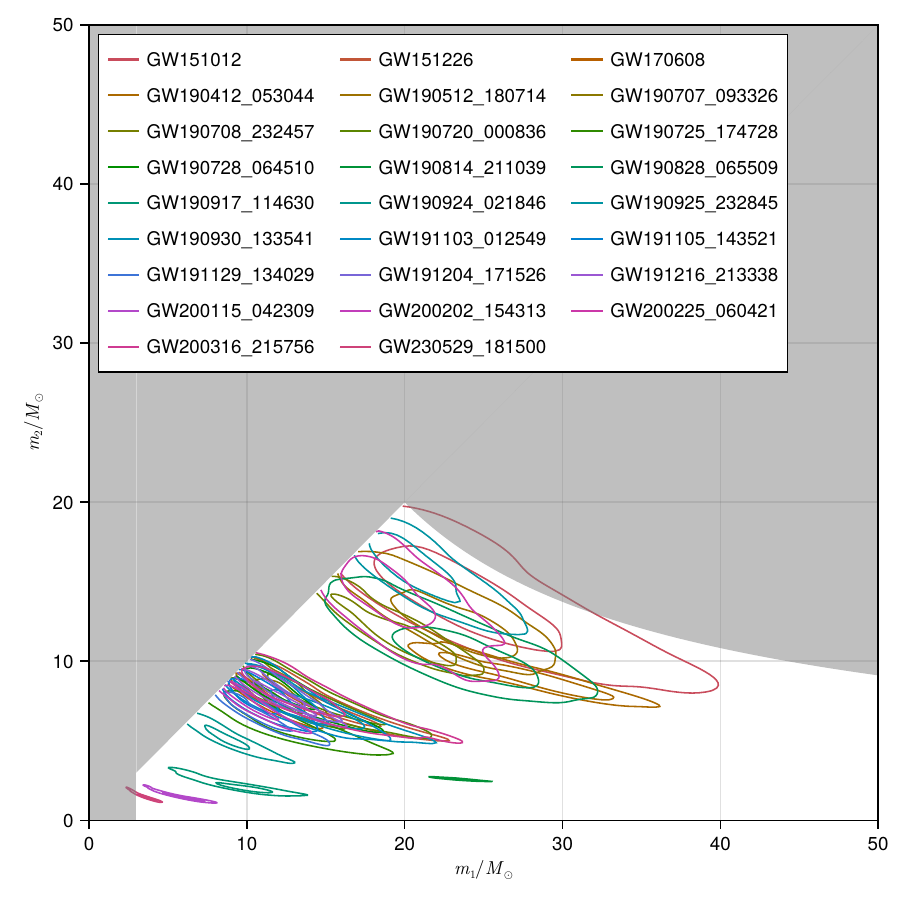}
    \caption{\label{fig:m1-m2-contour} Contour plot of the likelihood functions
    for the primary and secondary black hole masses in the events considered in
    this analysis.  The contours show credible regions containing $50\%$ and
    $90\%$ of the likelihood for each event.  The dashed lines show our
	selection cuts, with $m_1 > 3 \, M_\odot$ and $\mathcal{M} <
    \mchighunits$.}
\end{figure}

To restrict to low-mass events, we impose the requirement that $50\%$ of posterior samples
for each event lie within our chosen mass-cuts~\citep{Roy:2025ktr}.  Technically speaking, because
the posterior distribution sampled by parameter estimation is a function of the
data, these cuts satisfy the model assumptions of the selected population models
in \citet{Mandel:2018mve}; but to properly implement the normalization necessary
to correct for selection, we would need to perform similar parameter estimation
on the LVK sample of detected injections \citep{KAGRA:2021duu, Roy:2025ktr}, which is
extremely challenging.  Instead, we choose to implement our cuts in chirp mass,
a fairly well-measured variable, and treat the detected injections
asymmetrically to the catalog events, rejecting any injection whose true
parameter values lie outside our mass cut.  See Fig,~\ref{fig:m1-m2-contour} for
a visualization of our mass cuts in the $m_1-m_2$ parameter space and the
posterior distributions of the selected events; because chirp mass is so
well-measured, only one event's inclusion in the analysis set depends in any
meaningful way on the 50\% threshold chosen (i.e.\ for all but one event, our
mass cuts are equivalent to imposing the same cut on the---unknown---true
masses). During inference, we further impose
$(dN/dm_1dm_2dz)=0$ outside of the cuts and normalize the distributions
accordingly. 

In other words, we are inferring our mass function from a dataset of CBCs that
contain at least one BH (a compact object heavier than $m_\mathrm{low}$) and
constraining the minimum BH mass as a percentile of this distribution after
normalizing it in between $m_\mathrm{low}<m<m_\mathrm{high}$~(see
Eq.~\eqref{eq:pm-definition}). This justifies a direct comparison of our
measurements with LMXB studies~\citep{Bailyn:1997xt, Farr:2010tu, Ozel:2010su}
which also attempt to constrain the minimum BH mass from a dataset of observed
systems whose only compact object components are BHs, in contrast to other GW-based studies~\citep{Farah:2021qom}
which instead look for a dearth in the full compact object mass spectrum near
the maximum NS mass from a dataset of BNS, NSBH, and BBH systems.

\newpage
\section{Results}
\label{sec:results}
\begin{figure}
    \includegraphics[width=\columnwidth]{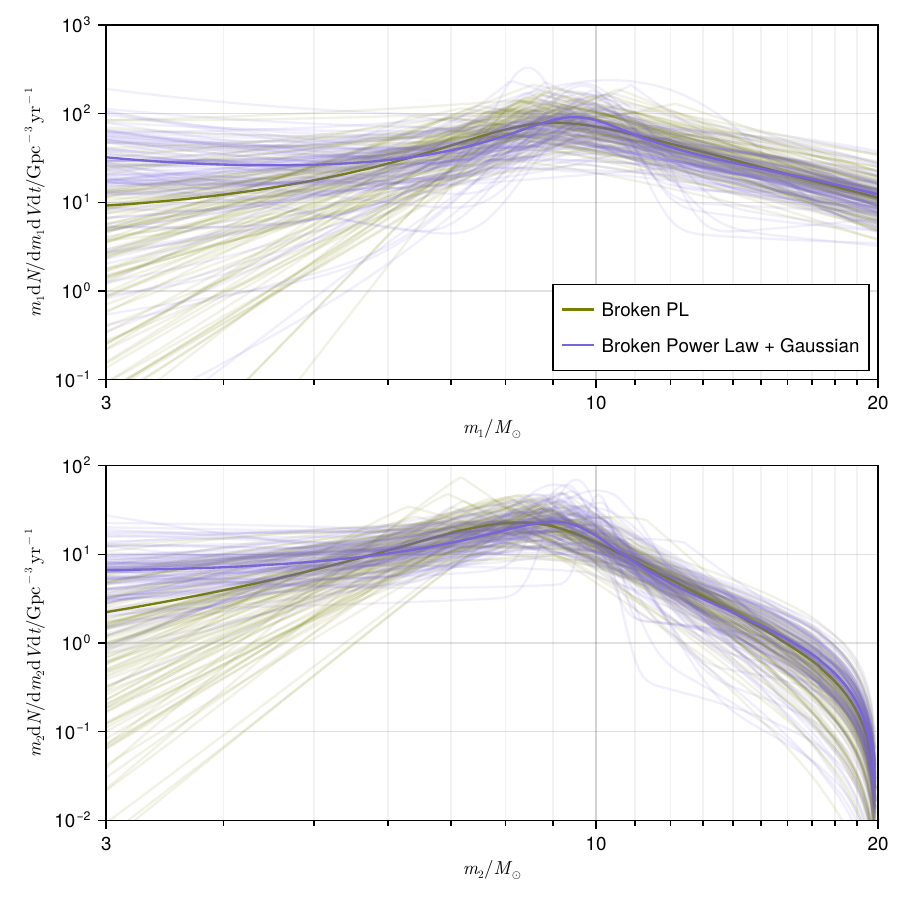}
    \caption{\label{fig:dNdm-traces_including_230529} Inferred mass functions for $3 \, M_\odot <
    m_2 < m_1 < 20 \, M_\odot$ from two models; both primary and secondary
    (marginal) mass functions are shown.  Dark lines show the posterior mean
    mass function; light lines are individual draws from the posterior over mass
    functions.}
\end{figure}

\begin{figure}
    \includegraphics[width=\columnwidth]{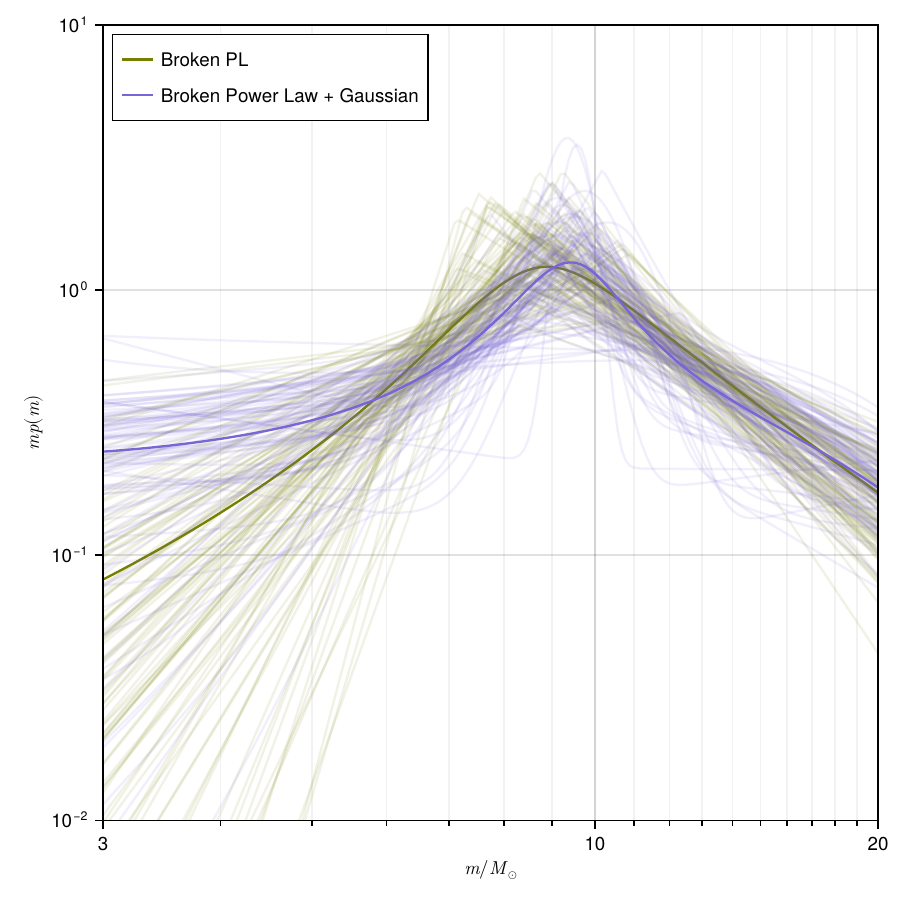}
    \caption{\label{fig:pm-traces_including_230529} Inferred common mass
    distribution, $p(m)$, (see Eq.~\eqref{eq:pm-definition}) for both models
    considered in this analysis.  Dark lines show the posterior mean mass
    distribution; light lines are individual draws from the posterior over mass
    distributions.  At high masses, $m \gg \mu, m_b$, the mass function falls
    steeply in both models; the broken power law plus Gaussian slope $\alpha_2 =
    \alphatwoplgplp$. Toward lower masses from the peak, both power law models
    initially decline significantly, though the power law plus Gaussian model
    may rise again as $m \to 3 \, M_\odot$.}
\end{figure}

In this section, we present our results obtained from the 25 GWTC-3 events that
satisfy our detection threshold~(i.e. a false alarm rate of less than one per
year) and mass-cuts along with the mass gap event GW230529\_181500. We use single-event
posterior samples publicly released by the LVK~\citep{gwosco3, gwosco4} for each
observation to construct our likelihood. We also use LVK's publicly released set
of detectable injections~\citep{gwosco3} to correct for selection biases in the
analysis \citep{Mandel:2018mve}. Note that this set of detectable simulations
estimates detector sensitivity through the LVK's third observation run which we
expect to suffice since GW230529\_181500 was detected within the first two weeks of the
fourth observing run~\citep{LIGOScientific:2024elc}.

In Fig.~\ref{fig:dNdm-traces_including_230529}, we show the inferred
distributions of individual component BH masses and In
Fig.~\ref{fig:pm-traces_including_230529}, the common BH mass function of
Eq.~\eqref{eq:pm-definition}. Using both of our models we find a strong excess
of BHs near $m=\mpeakplgplpincludingnewunits$, with an overall merger rate density of
\dNlogmpeakincludingnewunits per log mass squared, with the numbers corresponding to the BPLG model. The merger rate density is found to fall off by nearly an order of
magnitude on either side of this peak. We find that our models tend to disagree
towards the lower end of the mass spectrum with the BPL model predicting a
sharper drop in merger rate density. This behavior is expected given the lower
flexibility of the BPL model which tries to fit the peak as well as the
fall-offs using just two powerlaws. We however find consistency within
measurement uncertainties between the BPL and BPLG models.

\begin{figure}
    \includegraphics[width=\columnwidth]{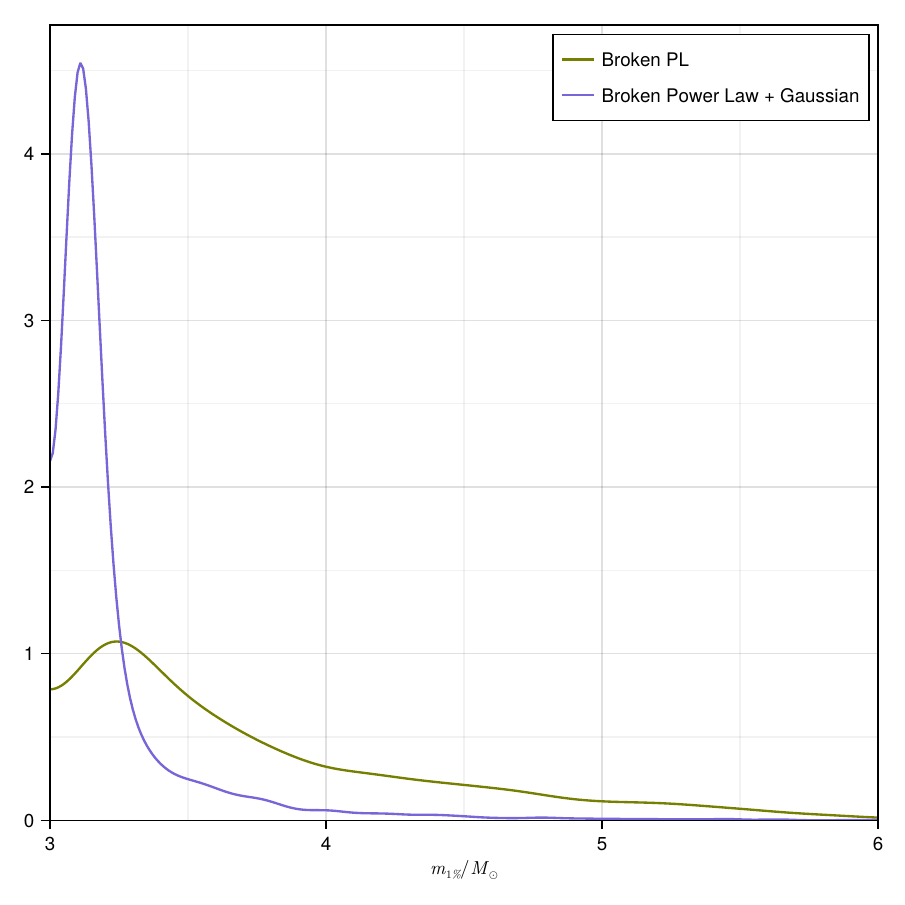}
    \caption{\label{fig:m1pct_including_230529} The posterior distribution for $m_{1\%}$, the
    first percentile of the ``common'' mass function for our three models.  The
    broken power law plus Gaussian model with 50\% selection cut has $m_{1\%} =
    \monepctplgplpincludingnewunits$ at $1\sigma$ (68\%) credibility.  The solid lines are
    the result for the 50\% selection cut, dashed for the 90\% selection cut.}
\end{figure}

In Fig.~\ref{fig:m1pct_including_230529}, we show our posterior distribution for the first percentile of BH masses which are found to peak near and have significant support at $m=m_\mathrm{low}$, and in table~\ref{tab:monepct} the corresponding 90\% highest posterior density credible intervals. The BPLG model yields a much tighter constraint on $m_{1\%}$ and has no posterior support for the minimum BH mass to be above $3.6M_{\odot}$. In other words, BPLG model rules out the existence of a lower mass gap in our sample of low mass BHs with 90\% posterior probability while the BPL model does not necessitate a gap but may permit one.

\begin{deluxetable}{llll}
\tablecolumns{3}
\tablecaption{\label{tab:monepct} $m_{1\%}$ for our various models and using different selection functions.}
\tablehead{\colhead{Mass Function Model} & \colhead{$m_{1\%} / M_\odot$ (90\%)} & \colhead{$m_{1\%} / M_\odot$ range (90\%)}}
\startdata
\texttt{Broken PL}& $3.24^{+1.52}_{-0.18}$& $\left[3.1, 4.8 \right]$\\ 
\texttt{Broken Power Law + Gaussian}& $3.10^{+0.39}_{-0.08}$& $\left[3.0, 3.5 \right]$\\ 
\enddata
\end{deluxetable}

\begin{figure}
     \includegraphics[width=\columnwidth]{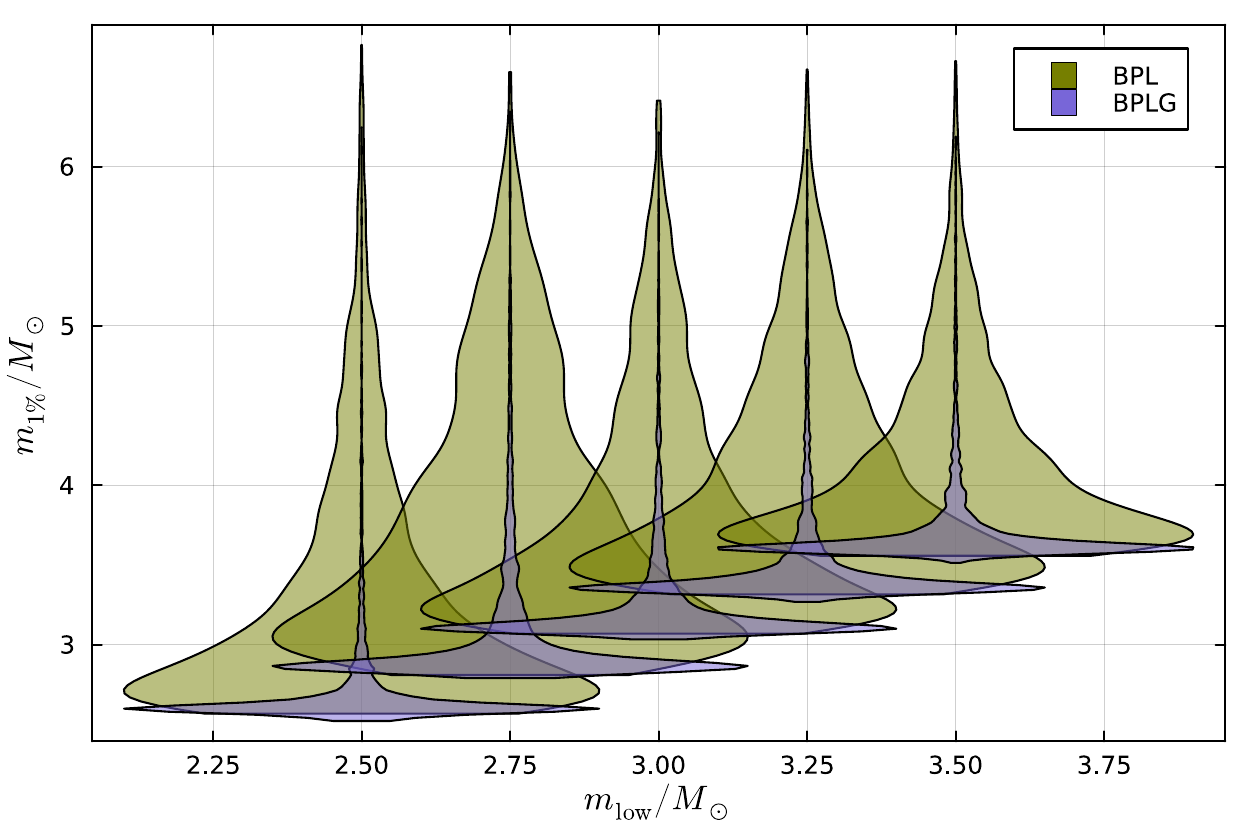}
     \caption{\label{fig:m1pct_including_230529_varying} The posterior for $m_{1\%}$, corresponding to different values of $m_{\mathrm{low}}$ for the BPLG model in the form of violin plots.}
 \end{figure}

To explore the existence of any systematic biases arising from an arbitrarily
chosen $m_\mathrm{low}$, which describes the boundary between the
neutron star and black hole mass functions in our models, we look at variations
in the measured $m_{1\%}$ values when the fixed $m_\mathrm{low}$ value is taken
to be something in the range $(2.5M_{\odot},3.5M_{\odot})$, which is shown in
Fig.~\ref{fig:m1pct_including_230529_varying}. We find that the peak of the
posterior closely follows the location of $m_\mathrm{low}$, which indicates the
absence of a lower mass-gap for all $m_\mathrm{low}$ values chosen. Both models
yield more constrained and peaked posteriors for smaller values of
$m_{\mathrm{low}}$. We see the posteriors start to grow broader with increasing
$m_\mathrm{low}$. The broadening of the posterior for $m_\mathrm{low}<3M_{\odot}$ as well
as the location of its peak is expected since in that range the secondary
component of GW190814~\citep{LIGOScientific:2020zkf} stops contributing to the
BH mass function for $m_\mathrm{low}>2.7M_{\odot}$, at which point
the $m_{1\%}$ inference is driven by the primary component of
GW230529\_181500. For  $m_\mathrm{low}>3.25M_{\odot}$, posterior samples of GW230529\_181500 start
getting excluded leading to weaker inference of the mass distribution in the
lower end since that is the only event that contributes dominantly to the
inference.

\section{Discussion}
\label{sec:discussion}
We have constrained the minimum BH mass using GW observations of CBC systems at least one of whose components is a BH. Using flexibly parametrized population models and a mass cut to filter out BNS and high mass BBH events, we find the posterior distribution of the first percentile of the BH mass function to be 90\% within \monepctplgplpincludingnewunits for our most flexible model (broken power-law with a Gaussian peak). Our methodology justifies a direct comparison with LMXB-based probes of the lower mass gap modulo their treatment of selection effects. We find that for our most flexible population model, the existing sample of GW detections does not exhibit a lower mass gap in contrast to LMXB-based measurements, which can indicate the latter to be Malmquist biased~\citep{Siegel:2022gwc}. For our less flexible model, however, we find that the current sample of GW detections does not require a gap but may permit one, with the $m_{1\%}$ posterior having small yet non-vanishing support near $5M_{\odot}$.

In addition to transient LMXB selection effects, an alternative explanation for this discrepancy is plausible. Merging BBH and NSBH systems with BHs in the lower mass gap can form through several scenarios such as hierarchical mergers in multiple systems~\citep{Lu:2020gfh, Liu:2020gif}, dynamical encounters in young metal-rich stellar clusters~\citep{ArcaSedda:2021zmm}, non-segregated clusters~\citep{Clausen:2014ksa, Fragione:2020wac, Rastello:2020sru}, and near active galactic nuclei~\citep{McKernan:2020lgr, Yang:2020xyi}, and massive stellar binaries~\citep{Antoniadis:2021dhe}, that are not expected to yield LMXB systems. Identifying the contributions of these specific evolutionary pathways to low mass NSBH and BBH formation might necessitate the inference of higher dimensional population distributions that model not only the mass spectrum but also its correlations with other CBC parameters such as component spin magnitudes and orientations. However, given the small number of events in the lower gap, such an inference is likely to be uninformative given current datasets and is hence left as a future exploration.

In accordance to existing studies~\citep[e.g.,][]{KAGRA:2021duu, LIGOScientific:2024elc, Tiwari:2021yvr, Farah:2023vsc, Ray:2023upk, Sadiq:2023zee, Edelman:2022ydv, Callister:2023tgi, Ray:2023upk, Ray:2024hos, Heinzel:2024hva}, we also recover an excess
of BHs near $m=\mpeakplgplpincludingnewunits$, with the merger rate density at the location of the being \dNlogmpeakincludingnewunits per log mass squared. We further characterize the fall-off in the merger rate density on either side of the peak. In particular, we find that in the low-mass end, the maximum fall-off compared to the peak is less than an order of magnitude, with the rate density at $3M_{\odot}$ being \dNlogmmlowincludingnewunits per log mass squared. We note that while various population models used in previous analyses have identified these features individually,  we are able to reconstruct both simultaneously using a single mass model, thereby showing that the data is consistent with a mass function that has a relatively strong peak near $10M_{\odot}$ and a very shallow fall-off in merger rate density towards the lower end of said peak, right up to the NS region of the parameter space.

The presence of both a strong peak and a shallow drop in merger rate density towards the lower end of the BH mass spectrum has interesting astrophysical implications. Several models of compact binary formation, such as isolated stellar binaries undergoing stable
mass transfer without a common envelope phase~\citep{vanSon:2022myr}, can explain this abundance of $8-10M_{\odot}$ BBHs or the shallow gap between NSs and BHs individually. For this particular formation channel, \cite{vanSon:2022myr} demonstrate that the astrophysical prescriptions that give rise to a strong peak near $10M_{\odot}$ also predict a very sharp fall of in merger rate density, which can drop by several orders of magnitude even above $4M_{\odot}$. On the other hand, the physical assumptions that are consistent with a very shallow drop in merger rate density predict no significant over-abundance of BBHs in the $8-10M_{\odot}$ range. In other words, any single channel modeled by  \cite{vanSon:2022myr} to explore the stable mass-transfer-only formation scenario cannot fully explain the shape of the BH mass function that we find in the lower end. If one were to assume that \cite{vanSon:2022myr} have exhaustively and accurately explored the stable-mass-transfer scenario, then our results could indicate substantial contributions from other formation channels, such as common-envelope evolution followed by stable mass transfer or dynamical encounters, to the observed population of low mass BHs.

\label{sec:conclusion}
\section{Acknowledgements}
\begin{acknowledgments}
    We thank Jeff Andrews for comments on an early version of this work. This work was initiated at the Aspen Center for Physics, which is supported by National Science Foundation grant PHY-2210452. This
research was also supported in part through the computational resources and
staff contributions provided for the Quest high-performance computing facility
at Northwestern University, which is jointly supported by the Office of the
Provost, the Office for Research, and Northwestern University Information
Technology.   Some computations in this work were, in part, run at facilities
supported by the Scientific Computing Core at the Flatiron Institute, a division
of the Simons Foundation.  A.R. was supported by the National Science
Foundation~(NSF) award PHY-2207945. V.K. was supported by the Gordon and Betty
Moore Foundation (grant awards GBMF8477 and GBMF12341), through a Guggenheim
Fellowship, and the D.I. Linzer Distinguished University Professorship fund.
W.F.~is partially supported by the Center for Computational Astrophysics, part
of the Flatiron Institute, a division of the Simons Foundation.  The authors are
grateful for computational resources provided by the LIGO Laboratory and
supported by NSF Grants No. PHY-0757058 and No. PHY0823459. V.K. and A.R.
gratefully acknowledge the support of the NSF-Simons AI-Institute for the Sky
(SkAI) via grants NSF AST-2421845 and Simons Foundation MPS-AI-00010513. This
material is based upon work supported by NSF’s LIGO Laboratory which is a major
facility fully funded by the National Science Foundation. This research has made
use of data obtained from the Gravitational Wave Open Science Center
(gwosc.org), a service of LIGO Laboratory, the LIGO Scientific Collaboration,
the Virgo Collaboration, and KAGRA. 
\end{acknowledgments}

\software{\texttt{Makie.jl} \citep{DanischKrumbiegel2021}; \texttt{Turing.jl} \citep{ge2018t}; \texttt{zenodo-get} \citep{Volgyes2020}}

\clearpage

\bibliography{Bump10MSun}

\begin{thebibliography}{}
\expandafter\ifx\csname natexlab\endcsname\relax\def\natexlab#1{#1}\fi
\providecommand{\url}[1]{\href{#1}{#1}}
\providecommand{\dodoi}[1]{doi:~\href{http://doi.org/#1}{\nolinkurl{#1}}}
\providecommand{\doeprint}[1]{\href{http://ascl.net/#1}{\nolinkurl{http://ascl.net/#1}}}
\providecommand{\doarXiv}[1]{\href{https://arxiv.org/abs/#1}{\nolinkurl{https://arxiv.org/abs/#1}}}

\bibitem[{Aasi {et~al.}(2015)}]{LIGOScientific:2014pky}
Aasi, J., {et~al.} 2015, Class. Quant. Grav., 32, 074001, \dodoi{10.1088/0264-9381/32/7/074001}

\bibitem[{Abac {et~al.}(2024)}]{LIGOScientific:2024elc}
Abac, A.~G., {et~al.} 2024, Astrophys. J. Lett., 970, L34, \dodoi{10.3847/2041-8213/ad5beb}

\bibitem[{Abbott {et~al.}(2020)}]{LIGOScientific:2020zkf}
Abbott, R., {et~al.} 2020, Astrophys. J. Lett., 896, L44, \dodoi{10.3847/2041-8213/ab960f}

\bibitem[{Abbott {et~al.}(2021)}]{LIGOScientific:2020ibl}
---. 2021, Phys. Rev. X, 11, 021053, \dodoi{10.1103/PhysRevX.11.021053}

\bibitem[{Abbott {et~al.}(2023{\natexlab{a}})}]{KAGRA:2021duu}
---. 2023{\natexlab{a}}, Phys. Rev. X, 13, 011048, \dodoi{10.1103/PhysRevX.13.011048}

\bibitem[{Abbott {et~al.}(2023{\natexlab{b}})}]{KAGRA:2021vkt}
---. 2023{\natexlab{b}}, Phys. Rev. X, 13, 041039, \dodoi{10.1103/PhysRevX.13.041039}

\bibitem[{Abbott {et~al.}(2023{\natexlab{c}})}]{gwosco3}
---. 2023{\natexlab{c}}, {Open data from the third observing run of LIGO, Virgo, KAGRA and GEO}.
\newblock \doarXiv{2302.03676}

\bibitem[{Acernese {et~al.}(2015)}]{VIRGO:2014yos}
Acernese, F., {et~al.} 2015, Class. Quant. Grav., 32, 024001, \dodoi{10.1088/0264-9381/32/2/024001}

\bibitem[{Ai {et~al.}(2019)Ai, Gao, \& Zhang}]{Ai:2019rre}
Ai, S., Gao, H., \& Zhang, B. 2019, \dodoi{10.3847/1538-4357/ab80bd}

\bibitem[{Akutsu {et~al.}(2021)}]{KAGRA:2020agh}
Akutsu, T., {et~al.} 2021, PTEP, 2021, 05A102, \dodoi{10.1093/ptep/ptab018}

\bibitem[{{Alsing} {et~al.}(2018){Alsing}, {Silva}, \& {Berti}}]{Alsing2018}
{Alsing}, J., {Silva}, H.~O., \& {Berti}, E. 2018, \mnras, 478, 1377, \dodoi{10.1093/mnras/sty1065}

\bibitem[{Antoniadis {et~al.}(2022)Antoniadis, Aguilera-Dena, Vigna-G\'omez, Kramer, Langer, M\"uller, Tauris, Wang, \& Xu}]{Antoniadis:2021dhe}
Antoniadis, J., Aguilera-Dena, D.~R., Vigna-G\'omez, A., {et~al.} 2022, Astron. Astrophys., 657, L6, \dodoi{10.1051/0004-6361/202142322}

\bibitem[{Arca~Sedda(2021)}]{ArcaSedda:2021zmm}
Arca~Sedda, M. 2021, Astrophys. J. Lett., 908, L38, \dodoi{10.3847/2041-8213/abdfcd}

\bibitem[{Bailyn {et~al.}(1998)Bailyn, Jain, Coppi, \& Orosz}]{Bailyn:1997xt}
Bailyn, C.~D., Jain, R.~K., Coppi, P., \& Orosz, J.~A. 1998, Astrophys. J., 499, 367, \dodoi{10.1086/305614}

\bibitem[{Belczynski {et~al.}(2012)Belczynski, Wiktorowicz, Fryer, Holz, \& Kalogera}]{Belczynski:2011bn}
Belczynski, K., Wiktorowicz, G., Fryer, C., Holz, D., \& Kalogera, V. 2012, Astrophys. J., 757, 91, \dodoi{10.1088/0004-637X/757/1/91}

\bibitem[{Callister \& Farr(2024)}]{Callister:2023tgi}
Callister, T.~A., \& Farr, W.~M. 2024, Phys. Rev. X, 14, 021005, \dodoi{10.1103/PhysRevX.14.021005}

\bibitem[{Clausen {et~al.}(2014)Clausen, Sigurdsson, \& Chernoff}]{Clausen:2014ksa}
Clausen, D., Sigurdsson, S., \& Chernoff, D.~F. 2014, Mon. Not. Roy. Astron. Soc., 442, 207, \dodoi{10.1093/mnras/stu871}

\bibitem[{Danisch \& Krumbiegel(2021)}]{DanischKrumbiegel2021}
Danisch, S., \& Krumbiegel, J. 2021, Journal of Open Source Software, 6, 3349, \dodoi{10.21105/joss.03349}

\bibitem[{Edelman {et~al.}(2023)Edelman, Farr, \& Doctor}]{Edelman:2022ydv}
Edelman, B., Farr, B., \& Doctor, Z. 2023, Astrophys. J., 946, 16, \dodoi{10.3847/1538-4357/acb5ed}

\bibitem[{Essick \& Farr(2022)}]{Pdet2-essick}
Essick, R., \& Farr, W. 2022, Precision Requirements for Monte Carlo Sums within Hierarchical Bayesian Inference.
\newblock \doarXiv{2204.00461}

\bibitem[{Ezquiaga \& Holz(2022)}]{Ezquiaga:2022zkx}
Ezquiaga, J.~M., \& Holz, D.~E. 2022, Phys. Rev. Lett., 129, 061102, \dodoi{10.1103/PhysRevLett.129.061102}

\bibitem[{Farah {et~al.}(2023)Farah, Edelman, Zevin, Fishbach, Ezquiaga, Farr, \& Holz}]{Farah:2023vsc}
Farah, A.~M., Edelman, B., Zevin, M., {et~al.} 2023, Astrophys. J., 955, 107, \dodoi{10.3847/1538-4357/aced02}

\bibitem[{Farah {et~al.}(2022)Farah, Fishbach, Essick, Holz, \& Galaudage}]{Farah:2021qom}
Farah, A.~M., Fishbach, M., Essick, R., Holz, D.~E., \& Galaudage, S. 2022, Astrophys. J., 931, 108, \dodoi{10.3847/1538-4357/ac5f03}

\bibitem[{Farr(2019)}]{Pdet1-Farr}
Farr, W.~M. 2019, Research Notes of the AAS, 3, 66, \dodoi{10.3847/2515-5172/ab1d5f}

\bibitem[{{Farr} \& {Chatziioannou}(2020)}]{Farr2020}
{Farr}, W.~M., \& {Chatziioannou}, K. 2020, Research Notes of the American Astronomical Society, 4, 65, \dodoi{10.3847/2515-5172/ab9088}

\bibitem[{Farr {et~al.}(2011)Farr, Sravan, Cantrell, Kreidberg, Bailyn, Mandel, \& Kalogera}]{Farr:2010tu}
Farr, W.~M., Sravan, N., Cantrell, A., {et~al.} 2011, Astrophys. J., 741, 103, \dodoi{10.1088/0004-637X/741/2/103}

\bibitem[{{Fishbach} \& {Holz}(2020)}]{Fishbach2020}
{Fishbach}, M., \& {Holz}, D.~E. 2020, \apjl, 891, L27, \dodoi{10.3847/2041-8213/ab7247}

\bibitem[{Fragione \& Banerjee(2020)}]{Fragione:2020wac}
Fragione, G., \& Banerjee, S. 2020, Astrophys. J. Lett., 901, L16, \dodoi{10.3847/2041-8213/abb671}

\bibitem[{Fryer {et~al.}(2012)Fryer, Belczynski, Wiktorowicz, Dominik, Kalogera, \& Holz}]{Fryer:2011cx}
Fryer, C.~L., Belczynski, K., Wiktorowicz, G., {et~al.} 2012, Astrophys. J., 749, 91, \dodoi{10.1088/0004-637X/749/1/91}

\bibitem[{Fryer \& Kalogera(2001)}]{Fryer:1999ht}
Fryer, C.~L., \& Kalogera, V. 2001, Astrophys. J., 554, 548, \dodoi{10.1086/321359}

\bibitem[{Fryer {et~al.}(2022)Fryer, Olejak, \& Belczynski}]{Fryer:2022lla}
Fryer, C.~L., Olejak, A., \& Belczynski, K. 2022, Astrophys. J., 931, 94, \dodoi{10.3847/1538-4357/ac6ac9}

\bibitem[{Ge {et~al.}(2018)Ge, Xu, \& Ghahramani}]{ge2018t}
Ge, H., Xu, K., \& Ghahramani, Z. 2018, in International Conference on Artificial Intelligence and Statistics, {AISTATS} 2018, 9-11 April 2018, Playa Blanca, Lanzarote, Canary Islands, Spain, 1682--1690.
\newblock \url{http://proceedings.mlr.press/v84/ge18b.html}

\bibitem[{Heinzel {et~al.}(2025)Heinzel, Mould, \& Vitale}]{Heinzel:2024hva}
Heinzel, J., Mould, M., \& Vitale, S. 2025, Phys. Rev. D, 111, L061305, \dodoi{10.1103/PhysRevD.111.L061305}

\bibitem[{Homan \& Gelman(2014)}]{HMC-NUTS}
Homan, M.~D., \& Gelman, A. 2014, J. Mach. Learn. Res., 15, 1593–1623

\bibitem[{Kalogera \& Baym(1996)}]{Kalogera:1996ci}
Kalogera, V., \& Baym, G. 1996, Astrophys. J. Lett., 470, L61, \dodoi{10.1086/310296}

\bibitem[{Kreidberg {et~al.}(2012)Kreidberg, Bailyn, Farr, \& Kalogera}]{Kreidberg:2012ud}
Kreidberg, L., Bailyn, C.~D., Farr, W.~M., \& Kalogera, V. 2012, Astrophys. J., 757, 36, \dodoi{10.1088/0004-637X/757/1/36}

\bibitem[{{LIGO Scientific Collaboration} {et~al.}(2024){LIGO Scientific Collaboration}, {KAGRA Collaboration}, \& {Virgo Collaboration}}]{gwosco4}
{LIGO Scientific Collaboration}, {KAGRA Collaboration}, \& {Virgo Collaboration}. 2024, LVK data release for GW230529\_181500 event,  Gravitational Wave Open Science Center, \dodoi{10.7935/6K89-7Q62}

\bibitem[{Liotine {et~al.}(2023)Liotine, Zevin, Berry, Doctor, \& Kalogera}]{Liotine2023}
Liotine, C., Zevin, M., Berry, C. P.~L., Doctor, Z., \& Kalogera, V. 2023, The Astrophysical Journal, 946, 4, \dodoi{10.3847/1538-4357/acb8b2}

\bibitem[{Liu \& Lai(2021)}]{Liu:2020gif}
Liu, B., \& Lai, D. 2021, Mon. Not. Roy. Astron. Soc., 502, 2049, \dodoi{10.1093/mnras/stab178}

\bibitem[{Liu {et~al.}(2021)Liu, Wei, Xue, \& Sun}]{Liu:2020uba}
Liu, T., Wei, Y.-F., Xue, L., \& Sun, M.-Y. 2021, Astrophys. J., 908, 106, \dodoi{10.3847/1538-4357/abd24e}

\bibitem[{Loredo(2004)}]{popgw2}
Loredo, T.~J. 2004, in {AIP} Conference Proceedings ({AIP}), \dodoi{10.1063/1.1835214}

\bibitem[{Lu {et~al.}(2020)Lu, Beniamini, \& Bonnerot}]{Lu:2020gfh}
Lu, W., Beniamini, P., \& Bonnerot, C. 2020, Mon. Not. Roy. Astron. Soc., 500, 1817, \dodoi{10.1093/mnras/staa3372}

\bibitem[{{Madau} \& {Dickinson}(2014)}]{Madau2014}
{Madau}, P., \& {Dickinson}, M. 2014, \araa, 52, 415, \dodoi{10.1146/annurev-astro-081811-125615}

\bibitem[{Mandel {et~al.}(2019)Mandel, Farr, \& Gair}]{Mandel:2018mve}
Mandel, I., Farr, W.~M., \& Gair, J.~R. 2019, Mon. Not. Roy. Astron. Soc., 486, 1086, \dodoi{10.1093/mnras/stz896}

\bibitem[{Mandel \& M\"uller(2020)}]{Mandel:2020qwb}
Mandel, I., \& M\"uller, B. 2020, Mon. Not. Roy. Astron. Soc., 499, 3214, \dodoi{10.1093/mnras/staa3043}

\bibitem[{Margalit \& Metzger(2017)}]{Margalit_2017}
Margalit, B., \& Metzger, B.~D. 2017, The Astrophysical Journal Letters, 850, L19, \dodoi{10.3847/2041-8213/aa991c}

\bibitem[{McKernan {et~al.}(2020)McKernan, Ford, \& O'Shaughnessy}]{McKernan:2020lgr}
McKernan, B., Ford, K. E.~S., \& O'Shaughnessy, R. 2020, Mon. Not. Roy. Astron. Soc., 498, 4088, \dodoi{10.1093/mnras/staa2681}

\bibitem[{Mueller \& Serot(1996)}]{Mueller:1996pm}
Mueller, H., \& Serot, B.~D. 1996, Nucl. Phys. A, 606, 508, \dodoi{10.1016/0375-9474(96)00187-X}

\bibitem[{Neal(2011)}]{HMC}
Neal, R.~M. 2011, in Handbook of Markov Chain Monte Carlo, ed. S.~Brooks, A.~Gelman, G.~Jones, \& X.-L. Meng (Chapman and Hall/{CRC}), \dodoi{10.1201/b10905}

\bibitem[{\"Ozel \& Freire(2016)}]{Ozel:2016oaf}
\"Ozel, F., \& Freire, P. 2016, Ann. Rev. Astron. Astrophys., 54, 401, \dodoi{10.1146/annurev-astro-081915-023322}

\bibitem[{Ozel {et~al.}(2010)Ozel, Psaltis, Narayan, \& McClintock}]{Ozel:2010su}
Ozel, F., Psaltis, D., Narayan, R., \& McClintock, J.~E. 2010, Astrophys. J., 725, 1918, \dodoi{10.1088/0004-637X/725/2/1918}

\bibitem[{Patton {et~al.}(2022)Patton, Sukhbold, \& Eldridge}]{Patton:2021gwh}
Patton, R.~A., Sukhbold, T., \& Eldridge, J.~J. 2022, Mon. Not. Roy. Astron. Soc., 511, 903, \dodoi{10.1093/mnras/stab3797}

\bibitem[{Raaijmakers {et~al.}(2021)Raaijmakers, Greif, Hebeler, Hinderer, Nissanke, Schwenk, Riley, Watts, Lattimer, \& Ho}]{Raaijmakers:2021uju}
Raaijmakers, G., Greif, S.~K., Hebeler, K., {et~al.} 2021, Astrophys. J. Lett., 918, L29, \dodoi{10.3847/2041-8213/ac089a}

\bibitem[{Rastello {et~al.}(2020)Rastello, Mapelli, Di~Carlo, Giacobbo, Santoliquido, Spera, Ballone, \& Iorio}]{Rastello:2020sru}
Rastello, S., Mapelli, M., Di~Carlo, U.~N., {et~al.} 2020, Mon. Not. Roy. Astron. Soc., 497, 1563, \dodoi{10.1093/mnras/staa2018}

\bibitem[{Ray {et~al.}(2024)Ray, Maga{\~n}a~Hernandez, Breivik, \& Creighton}]{Ray:2024hos}
Ray, A., Maga{\~n}a~Hernandez, I., Breivik, K., \& Creighton, J. 2024.
\newblock \doarXiv{2404.03166}

\bibitem[{Ray {et~al.}(2023)Ray, Maga{\~n}a~Hernandez, Mohite, Creighton, \& Kapadia}]{Ray:2023upk}
Ray, A., Maga{\~n}a~Hernandez, I., Mohite, S., Creighton, J., \& Kapadia, S. 2023, Astrophys. J., 957, 37, \dodoi{10.3847/1538-4357/acf452}

\bibitem[{Rhoades \& Ruffini(1974)}]{PhysRevLett.32.324}
Rhoades, C.~E., \& Ruffini, R. 1974, Phys. Rev. Lett., 32, 324, \dodoi{10.1103/PhysRevLett.32.324}

\bibitem[{Roy {et~al.}(2025)Roy, van Son, \& Farr}]{Roy:2025ktr}
Roy, S.~K., van Son, L. A.~C., \& Farr, W.~M. 2025.
\newblock \doarXiv{2507.01086}

\bibitem[{Sadiq {et~al.}(2024)Sadiq, Dent, \& Gieles}]{Sadiq:2023zee}
Sadiq, J., Dent, T., \& Gieles, M. 2024, Astrophys. J., 960, 65, \dodoi{10.3847/1538-4357/ad0ce6}

\bibitem[{Shao {et~al.}(2020)Shao, Tang, Jiang, \& Fan}]{Shao:2020bzt}
Shao, D.-S., Tang, S.-P., Jiang, J.-L., \& Fan, Y.-Z. 2020, Phys. Rev. D, 102, 063006, \dodoi{10.1103/PhysRevD.102.063006}

\bibitem[{Siegel {et~al.}(2023)}]{Siegel:2022gwc}
Siegel, J.~C., {et~al.} 2023, Astrophys. J., 954, 212, \dodoi{10.3847/1538-4357/ace9d9}

\bibitem[{{Talbot} \& {Golomb}(2023)}]{Talbot2023}
{Talbot}, C., \& {Golomb}, J. 2023, \mnras, 526, 3495, \dodoi{10.1093/mnras/stad2968}

\bibitem[{Thrane \& Talbot(2019)}]{Thrane:2018qnx}
Thrane, E., \& Talbot, C. 2019, Publ. Astron. Soc. Austral., 36, e010, \dodoi{10.1017/pasa.2019.2}

\bibitem[{Tiwari(2022)}]{Tiwari:2021yvr}
Tiwari, V. 2022, Astrophys. J., 928, 155, \dodoi{10.3847/1538-4357/ac589a}

\bibitem[{van Son {et~al.}(2022)van Son, de~Mink, Renzo, Justham, Zapartas, Breivik, Callister, Farr, \& Conroy}]{vanSon:2022myr}
van Son, L. A.~C., de~Mink, S.~E., Renzo, M., {et~al.} 2022, Astrophys. J., 940, 184, \dodoi{10.3847/1538-4357/ac9b0a}

\bibitem[{Vitale {et~al.}(2020)Vitale, Gerosa, Farr, \& Taylor}]{pop-vitale}
Vitale, S., Gerosa, D., Farr, W.~M., \& Taylor, S.~R. 2020, \dodoi{10.1007/978-981-15-4702-7_45-1}

\bibitem[{V\"{o}lgyes(2020)}]{Volgyes2020}
V\"{o}lgyes, D. 2020, Zenodo\_get: a downloader for Zenodo records., \dodoi{10.5281/zenodo.1261812}

\bibitem[{Wysocki {et~al.}(2019)Wysocki, Lange, \& O'Shaughnessy}]{popgw3}
Wysocki, D., Lange, J., \& O'Shaughnessy, R. 2019, Phys. Rev. D, 100, 043012, \dodoi{10.1103/PhysRevD.100.043012}

\bibitem[{Yang {et~al.}(2020)Yang, Gayathri, Bartos, Haiman, Safarzadeh, \& Tagawa}]{Yang:2020xyi}
Yang, Y., Gayathri, V., Bartos, I., {et~al.} 2020, Astrophys. J. Lett., 901, L34, \dodoi{10.3847/2041-8213/abb940}

\bibitem[{Zevin {et~al.}(2020)Zevin, Spera, Berry, \& Kalogera}]{Zevin:2020gma}
Zevin, M., Spera, M., Berry, C. P.~L., \& Kalogera, V. 2020, Astrophys. J. Lett., 899, L1, \dodoi{10.3847/2041-8213/aba74e}

\end{thebibliography}

\end{document}